\documentclass[aip,reprint]{revtex4-1} 
\usepackage[utf8]{inputenc}
\usepackage{amsfonts,amsmath,amssymb}
\usepackage{graphicx}
\usepackage{bm} 
\usepackage{mathrsfs}
\usepackage{subfigure}
\usepackage{textcomp}
\usepackage{hyperref}
\usepackage[flushleft]{threeparttable}
\usepackage[per-mode=symbol]{siunitx}
\usepackage[version=3]{mhchem}

\graphicspath{{../figures/}}

\DeclareSIUnit\intensity{\watt\per\centi\meter\squared}
\DeclareSIUnit\fieldstrength{\volt\per\centi\meter}
\usepackage{xspace}

\DeclareUnicodeCharacter{2009}{\,}

\newcommand{\cost}{\ensuremath{\langle\cos^2\theta_\text{2D}\rangle}}

\newlength{\figwidth}
\newlength{\figwidthwide}
\let\orgautoref\autoref
\providecommand{\Autoref}{%
  \def\equationautorefname{Equation}%
  \def\figureautorefname{Figure}%
  \def\subfigureautorefname{Figure}%
  \def\tableautorefname{Table}
  \def\sectionautorefname{Section}%
  \orgautoref}
\renewcommand{\autoref}{%
  \def\equationautorefname{Eq.}%
  \def\figureautorefname{Fig.}%
  \def\subfigureautorefname{Fig.}%
  \def\sectionautorefname{Sec.}%
  \def\tableautorefname{Tab.}
  \orgautoref}

\usepackage{color}
\usepackage[normalem]{ulem}
\definecolor{darkgreen}{rgb}{0.0,0.7,0.0}

\newcommand\singlet{$\mathrm{1^1\Sigma_g^+}$\xspace}
\newcommand\triplet{$\mathrm{1^3\Sigma_u^+}$\xspace}
\newcommand\Kdimer{$\mathrm{K_2}$\xspace}
\newcommand\Rbdimer{$\mathrm{Rb_2}$\xspace}
\newcommand\Nadimer{$\mathrm{Na_2}$\xspace}
\newcommand\Akion{$\mathrm{Ak^+}$\xspace}
\newcommand\Kion{$\mathrm{K^+}$\xspace}

\begin{document}

\preprint{draft 1}


\title{Time-resolved Coulomb explosion imaging of vibrational wave packets in alkali dimers on helium nanodroplets} 



\author{Nicolaj K. Jyde}
\altaffiliation{}
\affiliation{Department of Chemistry, Aarhus University, Langelandsgade 140, 8000 Aarhus C, Denmark}

\author{Henrik H. Kristensen}
\altaffiliation{}
\affiliation{Department of Physics and Astronomy, Aarhus University, Ny Munkegade 120, 8000 Aarhus C, Denmark}

\author{Lorenz Kranabetter}
\altaffiliation{}
\affiliation{Department of Chemistry, Aarhus University, Langelandsgade 140, 8000 Aarhus C, Denmark}

\author{Jeppe K. Christensen}
\altaffiliation{}
\affiliation{Department of Chemistry, Aarhus University, Langelandsgade 140, 8000 Aarhus C, Denmark}

\author{Emil Hansen}
\altaffiliation{}
\affiliation{Department of Physics and Astronomy, Aarhus University, Ny Munkegade 120, 8000 Aarhus C, Denmark}

\author{Mads B. Carlsen}
\altaffiliation{}
\affiliation{Department of Physics and Astronomy, Aarhus University, Ny Munkegade 120, 8000 Aarhus C, Denmark}

\author{Henrik Stapelfeldt}
\email[]{henriks@chem.au.dk}
\altaffiliation{}
\affiliation{Department of Chemistry, Aarhus University, Langelandsgade 140, 8000 Aarhus C, Denmark}


\date{\today}

\begin{abstract}
Vibrational wave packets are created in the lowest triplet state \triplet of \Kdimer and \Rbdimer residing on the surface of helium nanodroplets, through non-resonant stimulated impulsive Raman scattering induced by a moderately intense near-infrared laser pulse. A delayed, intense 50-fs laser pulse doubly ionizes the alkali dimers via multiphoton absorption and thereby causes them to Coulomb explode into a pair of alkali ions \Akion. From the kinetic energy distribution $P(E_\mathrm{kin})$ of the \Akion fragment ions, measured at a large number of delays, we determine the time-dependent internuclear distribution $P(R,t)$, which represents the modulus square of the wave packet within the accuracy of the experiment. For both \Kdimer and \Rbdimer, $P(R,t)$ exhibits a periodic oscillatory structure throughout the respective 300~ps and 100~ps observation times. The oscillatory structure is reflected in the time-dependent mean value of $R$, $\langle R \rangle(t)$. Fourier transformation of $\langle R \rangle(t)$ shows that the wave packets are composed mainly of the vibrational ground state and the first excited vibrational state, in agreement with numerical simulations. In the case of \Kdimer, the oscillations are observed for $300$~ps corresponding to more than $180$ vibrational periods with an amplitude that decreases gradually from 0.035~\AA{} to 0.020~\AA. Using time-resolved spectral analysis, we find that the decay time of the amplitude is~$\sim$~260~ps. The decrease is ascribed to the weak coupling between the vibrating dimers and the droplet.


\end{abstract}

\pacs{}

\maketitle 

\section{Introduction}

Vibration is the fundamental internal motion of molecules. With the advent of femtosecond laser pulses, it became possible to create vibrational wave packets, i.e., coherent superpositions of vibrational eigenstates, and thereby observe~\cite{bowman_femtosecond_1989,baumert_femtosecond_1991,fischer_femtosecond_1995} and, sometimes, manipulate vibrational motion~\cite{ohmori_wave-packet_2009,brif_control_2010,brinks_visualizing_2010,christensen_dynamic_2014} in real-time. Furthermore, with the introduction of ultrafast structure-sensitive methods, it has also become possible to measure the time evolution of internuclear distances and angles pertaining to the particular vibrational motion explored and, in some cases, even of the internuclear wave function. The main techniques employed in such studies are timed diffraction of sufficiently short (femtosecond) bursts of photons~\cite{minitti_imaging_2015,glownia_self-referenced_2016,chergui_photoinduced_2017,haldrup_ultrafast_2019} or electrons~\cite{ihee_direct_2001,yang_diffractive_2016,yang_imaging_2018}, laser-induced electron diffraction~\cite{wolter_ultrafast_2016,giovannini_new_2023}, timed Coulomb explosion imaging~\cite{stapelfeldt_time-resolved_1998,folmer_arresting_1998,chelkowski_measuring_2002,legare_laser_2006,bocharova_time-resolved_2011} and X-ray absorption~\cite{rupprecht_resolving_2023}. Most studies were conducted on gas-phase samples and examples on molecules where time-resolved structural-imaging methods have been employed to explore vibrational motion include \ce{H2+}~\cite{rudenko_real-time_2006} and \ce{D2+}~\cite{ergler_spatiotemporal_2006}, iodine~\cite{stapelfeldt_wave_1995}, iodomethane~\cite{malakar_time-resolved_2019}, substituted biphenyls~\cite{madsen_manipulating_2009,hansen_control_2012}, and ultralong-range Rydberg molecules~\cite{zou_observation_2023}.

Here, we explore the fundamental vibrational motion of one of the simplest molecular systems namely that of alkali atom dimers, however not as isolated systems but instead residing at the surface of helium nanodroplets. The first observations of vibrational wave packets in alkali dimers were reported for gas-phase samples of \ce{Na2} in the early 1990s by Baumert, Gerbert, and coworkers~\cite*{baumert_femtosecond_1991,baumert_femtosecond_1992}. About 10-15 years later, Schulz, Mudrich, Stienkemeier and coworkers demonstrated that vibrational wave packets in alkali dimers can also be created and probed when the dimers are attached to helium nanodroplets~\cite{claas_wave_2006,claas_wave_2007,mudrich_spectroscopy_2009,schlesinger_dissipative_2010, gruner_vibrational_2011}. The wave packets were formed both in electronically excited states through one-photon excitation as well as in the initially populated electronic state through resonant impulsive stimulated Raman scattering induced by the femtosecond pump pulse.

Those studies opened new opportunities in at least two different ways. First, unlike the case of gas-phase samples, alkali dimers on the surface of helium droplets are created in not just the \singlet ground state but also in the lowest-lying triplet state \triplet. This allowed investigations of wave packet dynamics in weakly bound triplet states. Second, the alkali dimers can interact with the nearby helium droplet, and thus spectroscopy permitted time-resolved exploration of how this interaction influences the vibrational motion. It was found that while the vibrational frequencies are almost identical to those of isolated molecules, the wave packets gradually loose coherence mainly due to vibrational relaxation. For dimers in triplet states, the decoherence time was on the order of a nanosecond~\cite{gruner_vibrational_2011}, long enough that the dimers could undergo hundreds of vibrations. In this work, we also explore vibrational wave packets in alkali dimers on the surface of helium droplets focusing on \Kdimer and \Rbdimer in the \triplet state. The main purpose of our studies is to show that timed Coulomb explosion imaging makes it possible to determine the distribution of bond lengths as a function of time. This method and results extend beyond previous works where wave packets were investigated by recording time-dependent ionization yields, which do not capture the molecular structure.

\section{Principle of timed Coulomb explosion imaging of vibrating alkali dimers}

\begin{figure}[h!]
    \centering
    \includegraphics[width=.48\textwidth]{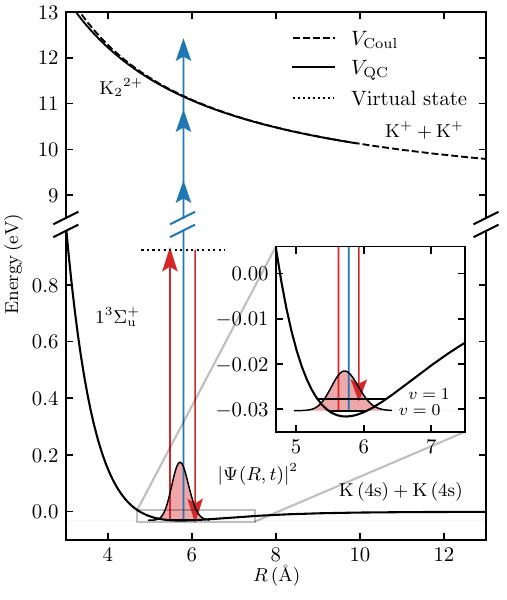}
    \caption{Energy diagram showing the potential curve for \ce{K2} in the \triplet state~\cite{bauer_accurate_2019} as well as the potential curves $V_\mathrm{QC}(R)$~\cite{kristensen_alignment_2024} and $V_\mathrm{Coul}(R)$ for ${\mathrm{K}_2}^{2+}$. The red vertical arrows represent photons from the pump laser pulse to illustrate the stimulated Raman process via a virtual state (dotted line), that creates the coherent superposition of the $v = 0$ and $v = 1$ vibrational eigenstates. The blue vertical arrows illustrate the multiphoton absorption process, induced by the probe pulse, which causes double ionization of \ce{K2} and subsequent Coulomb explosion into a pair of \ce{K+} ions. The red shaded area represents the square of the vibrational wave packet $\left| \Psi(R,t) \right|^2$. The inset is a close-up of the \triplet state potential curve.}
    \label{fig:cei}
\end{figure}

Using \Kdimer as an example, \autoref{fig:cei} schematically illustrates the pumping process that creates the vibrational wave packet and the probing process that measures the distribution of internuclear distances in the dimer at different times. When formed on the surface of helium nanodroplets, alkali dimers are populated in either the ground \singlet state or in the lowest-lying triplet state \triplet, the latter being the most abundant~\cite{higgins_helium_1998,bunermann_modeling_2011,kristensen_quantum-state-sensitive_2022}. \Autoref{fig:cei} shows the potential curve for the \triplet state.  The alkali dimers have been shown to equilibrate to the temperature $T = 0.37$~K at the droplet surface, similar to dopants located inside the droplets~\cite{aubock_triplet_2007}. At this temperature, only the $v = 0$ vibrational ground state is populated in the \triplet state of \Kdimer and \Rbdimer.

The central wavelength of the $600$-fs-long pump pulse is set to 1.30~$\mu$m  and therefore the photon energy is too low to electronically excite the \Kdimer~\cite{magnier_theoretical_2004} via single-photon absorption. This is also the case for \Rbdimer~\cite{allouche_transition_2012}. The intensity is, however, chosen sufficiently high such that some of the population in the initial $v = 0$ state can be transferred to the $v = 1$ state by a stimulated Raman transition, and potentially to higher-lying vibrational states. This method of preparing vibrational wave packets is well-established and has been thoroughly investigated, both experimentally and theoretically~\cite{ruhman_coherent_1988,madsen_manipulating_2009,thomas_non-resonant_2016,shu_femtochemistry_2017,rupprecht_resolving_2023}. In \autoref{sec:dse}, we provide an equivalent description of the excitation scheme as a transient distortion of the potential curve for the alkali dimer in order to present numerical simulations of the wave packet dynamics.

One advantage of stimulated Raman excitation is its efficiency, i.e., it ensures the formation of wave packets in essentially all molecules in the focal volume of the pump pulse. This is not the case when vibrational wave packets are created in electronically excited states through one-photon absorption. Here the intensity of the pump pulse must be kept sufficiently low to prevent two-photon excitation to higher-lying electronic states or even ionization. Consequently, a significant proportion of the molecules remain in their initial ground state. This leads to a background signal during the probing process, regardless of whether it involves X-ray diffraction, electron diffraction, or Coulomb explosion, as in our study. This is because the probing process typically interacts with any molecule, irrespective of its electronic state. Therefore, it is essential to apply a procedure for subtracting the background signal~\cite{stapelfeldt_time-resolved_1998,yang_diffractive_2016}. In the current study, we avoid background subtraction thanks to the Raman excitation scheme.

To measure the wave packet, we use timed Coulomb explosion, triggered by the probe pulse sent at time $t$ with respect to the center of the pump pulse. The 50 fs probe pulse doubly ionizes \ce{K2} via multiphoton absorption, illustrated by the blue arrows in \autoref{fig:cei}, and therefore projects the wave packet onto the repulsive potential curve of the dication \ce{K2^{2+}}. Since the two (i.e., all) valence electrons are removed from the dimer upon double ionization, the resulting closed-shell configuration of the \ce{K2^{2+}} state exhibits only a single, repulsive potential, yielding a one-to-one correspondence between the initial internuclear distance $R$ of \Kdimer and the potential curve $V_\text{dicat}(R)$ of \ce{K2^{2+}}.

Due to the repulsive character of $V_\text{dicat}$, \ce{K2^{2+}} breaks apart into two \Kion ions, the process referred to as Coulomb explosion. If \Kdimer is a homodimer, e.g. \ce{^39K2}, the final kinetic energy $E_{\text{kin}}$ of each \Kion ion acquired from Coulomb explosion from the equilibrium distance $R_\text{eq}$ is $\frac{1}{2} ( V_\text{dicat}(R_\text{eq})-2I_\text{p}(\text{K}) )$ where $I_\text{p}(\text{K})$ denotes the ionization energy of K~\footnote{If \ce{K2} is a heterodimer, e.g. \ce{^{39}K^{41}K}, the \ce{^{39}K+} and \ce{^{41}K+} ions acquire energy according to momentum conservation}. At the internuclear distances of \ce{Ak2} in the \triplet state pertinent to this work, $V_\text{dicat}$ is to a very good approximation given by a Coulomb potential $V_\text{Coul}$~\footnote{For instance, the \ce{K2^{2+}} potential curve was recently calculated at the CCSD(T) level of theory. The result, shown as $V_\mathrm{QC}(R)$ in Fig. 1, is essentially identical to $V_\text{Coul}$ for $R\geq$ 4.5~\AA.} and therefore:
\begin{equation}
\label{eq:Ekin}
\begin{aligned}
E_\text{kin} \ = \dfrac{7.2~\text{eV}}{R\text{ [\AA]}} \, .
\end{aligned}
\end{equation}
In the experiment, we record the distribution of kinetic energies $P(E_{\text{kin}})$ at all times. Thanks to \autoref{eq:Ekin}, we can then determine the distribution of $R$, $P(R)$, by a standard probability distribution transformation of $P(E_{\text{kin}})$ using the appropriate Jacobian~\cite{kristensen_laser-induced_2023}. Using this method, we retrieve the time-dependent internuclear distance distribution $P(R,t)$ which, within the uncertainty of the measurements, represents the experimentally determined square of the wave function $|\Psi(R,t)|^2$ of the vibrating alkali dimer. This approach was previously used to convert the experimental $P(E_{\text{kin}})$ into $P(R)$ for static alkali homonuclear and heteronuclear dimers, i.e. \ce{Ak2} and AkAk$^{\prime}$ in $v = 0$~\cite{kristensen_laser-induced_2023,albrechtsen_laser-induced_2024} and also for static alkali trimers~\cite{kranabetter_structure_2024}. Here we apply this procedure for a large number of delays between the pump and the probe pulses to obtain the time-dependent $R$-distribution for the vibrational wave packet.

\section{Dynamic Stark Effect simulation}
\label{sec:dse}
Non-resonant stimulated Raman scattering induced by a moderately intense laser pulse enables control of both the vibrational and rotational motion of molecules~\cite{townsend_stark_2011}. The interaction is described in terms of the polarizability interaction, and when averaged over a cycle of the laser field, it depends on the intensity of the laser pulse. In the case of a linear molecule interacting with a non-resonant, linearly polarized laser pulse, the dynamic Stark effect (DSE) is given by~\cite{sussman_quantum_2006,shu_femtochemistry_2017}:
\begin{align}\label{eq:pol_int_full}
V_\mathrm{DSE}(R,\theta,t) = &-\frac{\mathcal{E}_\mathrm{envl}^2(t)}{4} \{[\alpha_\parallel (R) - \alpha_\perp (R)] \cos^2 (\theta) \nonumber \\
&+ \alpha_\perp (R)\} \, ,
\end{align}
where $\mathcal{E}_\mathrm{envl}^2(t)$ is the squared field envelope function, and $\alpha_\parallel(R)$ and $\alpha_\perp(R)$ are the parallel and perpendicular components of the polarizability tensor, respectively. The control over the molecular dynamics is exerted through the dependence of \autoref{eq:pol_int_full} on the coordinates $R$ and $\theta$, where the former is the internuclear separation and the latter the angle between the internuclear axis and the polarization direction of the laser field. In particular, the $\theta$-dependence of $V_\mathrm{DSE}$ has been used to control the alignment of molecules, explored in a vast number of works over the past 25 years~\cite{stapelfeldt_colloquium:_2003,ohshima_coherent_2010,fleischer_molecular_2012,koch_quantum_2019}. Here we focus on the $R$-dependence of $V_\mathrm{DSE}$, which results from the $R$-dependence of the polarizability components, and how it can be used to initiate vibrational motion.


We examine the special case of a molecule aligned with the polarization axis of the laser field ($\theta = 0$). In this scenario, we can interpret the dynamic Stark effect as a transient modulation of the field-free potential curve $V_0(R)$, as expressed by the time-dependent potential~\cite{thomas_non-resonant_2016}
\begin{equation}
    \hat{V}(R,t) = \hat{V}_0(R) -\frac{{\mathcal{E}^2_{\mathrm{envl}}}(t)}{4}\alpha_\parallel(R) \, .
    \label{eq:V_int}
\end{equation}

\begin{figure}
    \centering
    \includegraphics{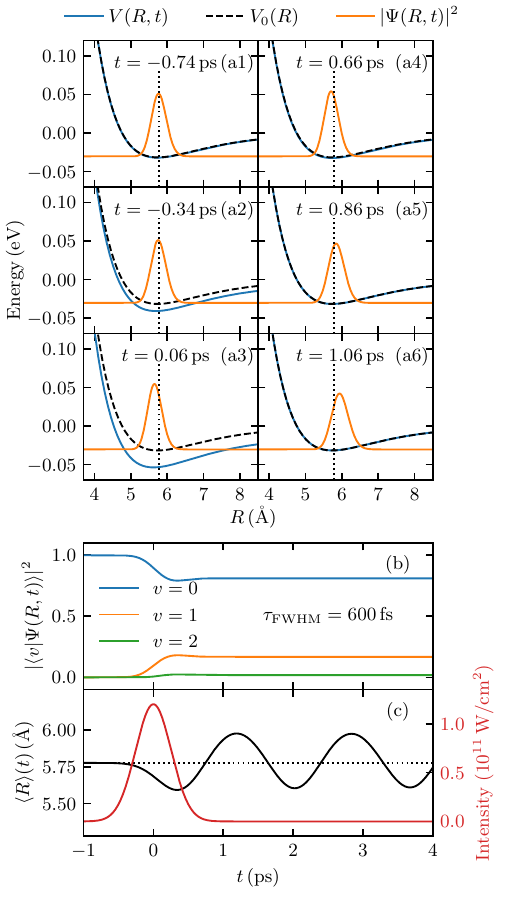}
    \caption{(a1)-(a6) Calculated $\left | \Psi(R,t) \right |^2$ (orange curve), time-dependent potential $V(R,t)$ (blue curve) and field-free potential $V_0(R)$ (dashed black curve) of $\mathrm{K}_2$ in the \triplet state, shown at different times before, during and after the interaction with the linearly polarized, $600$-fs Gaussian pump laser pulse centered at $t=0$. In all panels, the vertical dotted line shows $\langle R \rangle_{v=0}$. (b) The time-dependent populations for the three lowest vibrational eigenstates.
    (c) The intensity of the pump pulse (red curve) and $\langle R \rangle (t)$. (black curve). The horizontal dotted line shows $\langle R \rangle_{v=0}$.}
    \label{fig:dse}
\end{figure}

To illustrate the mechanism by which the vibrational wave packets are created and to enable a qualitative comparison with the experimental data, we present a numerical simulation of the dynamic Stark effect. To this end, we solved the one-dimensional, time-dependent Schrödinger equation with the vibrational Hamiltonian
\begin{equation}
    \hat{\mathrm{H}}=\hat{\mathrm{T}}+\hat{V}(R,t) \, ,
    \label{eq:hamiltonian}
\end{equation}
of an isolated \Kdimer molecule in the \triplet state, using the split-step operator method~\cite{fleck_time-dependent_1976}.
Here $\hat{\mathrm{T}}$ is the kinetic energy operator and we assume that the envelope of the square of the electric field is Gaussian and given by ${\mathcal{E}^2_\mathrm{envl}}(t)={\mathcal{E}^2_0} \exp\left [-4\ln(2)\left( t/\tau_{\mathrm{FWHM}}\right )^2 \right]$. The polarizability function $\alpha_\parallel (R)$ and the potential $V_0(R)$, which are used in \autoref{eq:V_int}, are taken from the literature~\cite{deiglmayr_calculations_2008,bauer_accurate_2019}. The initial state of the system is the vibrational ground state $|v=0\rangle$.



\Autoref{fig:dse}(a1)--(a6) show $\left| \Psi(R,t) \right|^2$ (orange curve) and $V(R,t)$ (blue curve) at six times covering just before, during and shortly after the interaction with the laser pulse. The six times are selected to coincide with those of the experimental data (\autoref{fig:wp}). The results are obtained for a pulse with $\tau_{\mathrm{FWHM}} = 600$~fs and peak intensity $I_\mathrm{peak}=1.2\times 10^{11} \, \mathrm{W/cm^2}$. At $t$ = $-$0.74~ps, the intensity of the laser pulse is still so low that $V(R,t)$ is essentially identical to $V_0(R)$ and, thus, $\left| \Psi(R,t) \right|^2$ is identical to that of the initial $v$ = 0 state. The expectation value of $R$ for this state, $\langle R \rangle_{v=0}$ = 5.77~\AA, is shown by the vertical dotted line. At $t$ = $-$0.34~ps, the laser pulse intensity is now so large that the minimum position of $V(R,t)$ has shifted to a smaller value compared to its field-free value (approximately equal to the position of the black vertical dotted line). This shift leads to an internuclear force, which drives the bond length of \Kdimer towards smaller values, visible as a minor shift of $\left| \Psi(R,t) \right|^2$ to the left. At $t$ = 0.06~ps, i.e., essentially when the laser intensity reaches its peak, the potential is further shifted and deepened and, therefore, the shortening of the bond distance continues, seen as a further shift of $\left| \Psi(R,t) \right|^2$ to the left. When the laser pulse intensity subsequently decreases, $V(R,t)$ gradually returns towards $V_0(R)$, and therefore \Kdimer is now exposed to an internuclear force driving the bond length to a larger value. As a result, $\left| \Psi(R,t) \right|^2$ is almost back to its initial position and shape at $t$ = 0.66~ps where the laser pulse is close to being completely turned off and $V(R,t)$ coincides with $V_0(R)$. The internuclear velocity is, however, not zero due to the internuclear excursion that was driven by the time-dependent potential. Therefore, the internuclear motion will continue at longer times in a periodic manner with an amplitude determined by the intensity and duration of the laser pulse. Panels (a5) and (a6) depict $\left| \Psi(R,t) \right|^2$ at $t$ = 0.66~ps and 0.86~ps, respectively, i.e., in the interval where the bond distance increases towards its maximum value.

More insight into the vibrational dynamics comes from plotting the population of the vibrational eigenstates as a function of time, obtained as $\left | \langle v | \Psi(R,t) \rangle \right |^2$.  This is done in \autoref{fig:dse}(b), which shows that the polarizability interaction transfers population from the initial $v$~=~0 state to the $v$~=~1 state and also a little to the $v$~=~2 state during the laser pulse. After the pulse, the population in the $v$~=~0, 1, and 2 states are 0.81, 0.17, and 0.02, respectively. As such, the polarizability interaction creates a coherent superposition of these three vibrational quantum states, i.e., a vibrational wave packet, $\Psi(t) = \sum_{v=0}^{2} c_v(t) |v\rangle e^{-\mathrm{i}E_vt/h}$. Consequently, the expectation value of $R$, $\langle R \rangle(t)=\langle \Psi(R,t) |R| \Psi(R,t) \rangle$ becomes time-dependent. \Autoref{fig:dse}(c), depicting $\langle R \rangle(t)$ for the first 4~ps after the center of the laser pulse, illustrates that the expectation value of $R$ oscillates periodically around $\langle R \rangle_{v=0}$ once the wave packet is formed. The sinusoidal shape results from the dominant coupling between the $v$~=~0 and the $v$~=~1 states, which according to standard analysis of wave packets has a frequency given by the energy difference between the two vibrational states,
\begin{equation}
    \nu_{1,0}=\frac{E_{1}-E_0}{h} \ .
    \label{eq:freqs}
\end{equation}
Analysis of $\langle R \rangle(t)$ at longer times reveals a tiny contribution from a periodic oscillation with a frequency corresponding to the energy difference between the $v$~=~1 and $v$~=~2 states. As discussed below, the experimental observations are consistent with these simulated results.

Concerning the choice of the duration and the intensity of the pump laser pulse, used in the simulations and the experiments, we note the following. Firstly, to create the wave packets, the pump pulse must transfer population from $v$~=~0 to excited vibrational states, but it should not lead to population of electronically excited states or ionization as a result of multiphoton absorption. Thus, for a given pulse duration, we choose an intensity that is as high as possible to maximize population of excited vibrational states yet low enough to avoid multiphoton electronic excitation and ionization. Secondly, to ensure a nonadiabatic interaction and thereby an efficient wave packet generation, the pulse duration should be significantly less than the vibrational period of the dimer. For \ce{K2}, the vibrational period, given by $(\nu_{{1,0}})^{-1}$, is 1.6 ps, which made us choose a pulse duration $\tau_{\mathrm{FWHM}} = 600$~fs. The intensity, $I_\mathrm{peak}=1.2 \times 10^{11} \, \mathrm{W/cm^2}$, was found experimentally by identifying the value where ionization becomes negligible. Likewise, for \ce{Rb2}, we chose $\tau_{\mathrm{FWHM}} = 900$~fs and $I_\mathrm{peak}=7.8 \times 10^{10} \, \mathrm{W/cm^2}$. For both \ce{K2} and \ce{Rb2} it is possible to induce wave packets with shorter pump pulses but the final populations in the vibrationally excited states become smaller.

\section{Experimental setup}
\label{sec-exp-setup}
A detailed description of the experimental setup has been given previously~\cite{kristensen_laser-induced_2023}, thus only the most important features of the setup are presented here.
A continuous helium droplet beam is generated by expanding high-purity helium-4 gas into vacuum through a cooled $5~\mu$m nozzle at a stagnation pressure of $25$~bar. The average droplet size depends on the stagnation pressure and nozzle temperature $T_\mathrm{nozzle}$. In the present work $T_\mathrm{nozzle}$~=~12~K which yields droplets with a mean size of $N_\mathrm{He}\sim$~10$^4$~He~atoms~\cite{toennies_superfluid_2004}. The droplet beam is skimmed before passing through a pickup cell containing a gas of either Rb or K atoms. The vapor pressure is adjusted such that the pick-up statistics favor the attachment of an average of two alkali atoms, yielding droplets doped with alkali dimers, which form in the lowest triplet state \triplet or in the ground state \singlet~\cite{stienkemeier_laser_1995, higgins_helium_1998, bruhl_triplet_2001, mudrich_formation_2004, lackner_spectroscopy_2013, kristensen_quantum-state-sensitive_2022}.
The doped helium droplet beam is subsequently skimmed and enters a velocity map imaging (VMI) spectrometer~\cite{chandler_twodimensional_1987,eppink_velocity_1997} in the center of which the droplet beam is crossed by two focused, pulsed 1~kHz laser beams. The pump pulses ($\lambda=1.30~\mu$m, \Kdimer: $I_\mathrm{peak}=1.2 \times 10^{11} \, \mathrm{W/cm^2}$ and $\tau_\mathrm{FWHM}=600$~fs, \Rbdimer: $I_\mathrm{peak}=7.8 \times 10^{10} \, \mathrm{W/cm^2}$ and $\tau_\mathrm{FWHM}=900$~fs) induce vibrational wave packets in the alkali dimers. The intensity of the pump pulses is limited to avoid any significant ionization of the dimers. The probe pulses ($\lambda=800$~nm, $\tau_\mathrm{FWHM}=50$~fs, \Kdimer: $I_\mathrm{peak}=2.2 \times 10^{13} \, \mathrm{W/cm^2}$, \Rbdimer: $I_\mathrm{peak}=6.7 \times 10^{13}\, \mathrm{W/cm^2}$ ) initiate Coulomb explosion of the dimers into a pair of $\mathrm{Ak}^+$ ions via multiphoton absorption. Finally, the 2D projection of the velocity of the fragment ions $E_\mathrm{kin}$ is recorded by a 2D imaging detector. The detector is gated such that only the ions with the desired mass-over-charge ratio are recorded. The positions of the ion hits are recorded by a CCD camera, where one frame contains ion hits accumulated from 10 laser pulses. Both laser beams are linearly polarized in the same direction parallel to the ion detector.

\section{Results and discussion}
\subsection{Ion images, kinetic energy distribution and internuclear distance distribution}
\begin{figure}[h!]
    \centering
    \includegraphics[width=.47\textwidth]{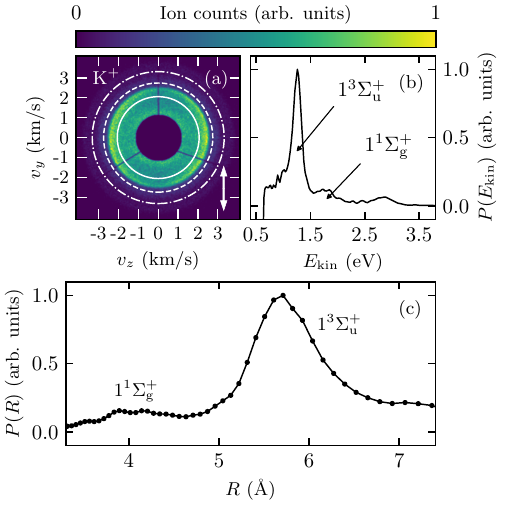}
    \caption{(a) Two-dimensional velocity image of \ce{^39K+} ions recorded with the probe pulse only. Its polarization axis is indicated by the vertical arrow. The white circles demarcate the radially separated Coulomb explosion channels pertaining to the \triplet (innermost channel) and \singlet (outermost channel) state, respectively. (b) Kinetic energy spectrum of $\mathrm{K^+}$ ions derived from the image in (a). The channels pertaining to the electronic states of the $\mathrm{K_2}$ dimer are annotated with their respective term symbols. (c) Distribution of internuclear distances $P(R)$ obtained by application of the Jacobian transformation of $P(E_\mathrm{kin})$ in (b).}
    \label{fig:data_analysis}
\end{figure}

\Autoref{fig:data_analysis}(a) shows a 2D velocity image of \ce{^39K+} ions obtained when helium droplets doped with potassium are irradiated with the probe pulses only. A dark circular feature at the center of the image, along with three radial lines, can be attributed to a metal disk and its supports~\cite{schouder_laser-induced_2020, chatterley_laser-induced_2020}. The disk is positioned in the flight tube, in front of the detector. The purpose of the disk is to eliminate detection of a large number of low-velocity alkali ions that originate from ionization of Ak atoms that have diffused into the chamber housing the VMI spectrometer or from ionization of Ak atoms on He droplet that picked up only a single atom. Additionally, the disk removes most \Akion ions resulting from the dissociative ionization of dimers.
Two distinct, radially separated channels are visible in the image, with their boundaries marked by white concentric circles.
As shown before~\cite{kristensen_quantum-state-sensitive_2022,kristensen_laser-induced_2023}, the \Kion hits between the solid and the dashed circles originate from Coulomb explosion of \Kdimer in the \triplet state, which is the system of interest here, while the hits between the dashed and dotted-dashed circle stem from Coulomb explosion of \Kdimer in the \singlet state~\cite{kristensen_quantum-state-sensitive_2022}.

To determine the kinetic energy distribution of the $\mathrm{K^+}$ ions from the 2D velocity images, we apply an Abel-inversion to the images. For this purpose, we utilize the Polar Onion Peeling (POP) algorithm~\cite{roberts_toward_2009}. The polarization axes of the probe pulses is parallel to the detector plane as required for the Abel inversion. From the Abel-inverted image, we calculate the radial distribution $P(r)$ by integrating over the angle. The kinetic energy distribution $P(E_\mathrm{kin})$ shown in \autoref{fig:data_analysis}(b) is related to $P(r)$ by the coordinate transformation $P(E_\mathrm{kin})= P(r) \left |\frac{\text{d}r}{\text{d}E_\mathrm{kin}(r)} \right|$, with $E_\mathrm{kin}(r)=kr^2$. Here, the constant of proportionality $k$ is chosen such that the best agreement is obtained with previous works~\cite{kristensen_quantum-state-sensitive_2022, kristensen_laser-induced_2023}. Finally, we obtain the distribution of internuclear separations $P(R)$, shown in \autoref{fig:data_analysis}(c), by using the Jacobian transformation $P(R)= P(E_\mathrm{kin}) \left |\frac{\text{d}E_\mathrm{kin}(R)}{\text{d}R} \right|$, where $E_\mathrm{kin}(R)$ is given by \autoref{eq:Ekin}. The region of interest is the part from $\sim$~4.8~\AA{} to $\sim$~7.0~\AA{} which represents $P(R)$ for the vibrational ground level in the \triplet state. The distribution is essentially identical to the one reported in Ref.~\cite{kristensen_laser-induced_2023}.


\subsection{Vibrational dynamics: $P(R,t)$ and $\langle R \rangle (t)$}
\begin{figure}[h!]
    \centering
    \includegraphics[width=.47\textwidth]{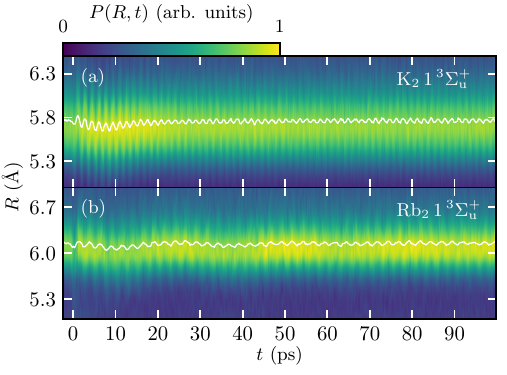}
    \caption{The experimental time-dependent distribution of internuclear separations, $P(R,t)$, for wave packets induced in the \triplet state of (a) $\mathrm{K_2}$ and (b) of $\mathrm{Rb_2}$. For each distribution the corresponding expectation value, $\langle R \rangle (t)$, is plotted as a white curve.}
    \label{fig:Ak2_wp_im}
\end{figure}

To measure the vibrational dynamics of the alkali dimers, the pump pulse was included (same polarization direction as the probe pulse), and we recorded velocity images of \ce{^39K+} and \ce{^85Rb+} ions for many delays between the pump and the probe pulses. For \Kdimer (\Rbdimer), images were recorded from $-2$~ps to $300$~ps in steps of $0.2$~ps ($-2$~ps to $100$~ps in steps of $0.25$~ps). These measurements enable the determination of the time-dependent probability distribution of the internuclear separation $P(R,t)$ of vibrational wave packets in the \triplet state of the two alkali dimers. The results, depicted in \autoref{fig:Ak2_wp_im}, show an oscillatory structure for both $\mathrm{K_2}$ and $\mathrm{Rb_2}$. To render the oscillations more clear, we determine the average value of $R$, $\langle R \rangle$, for each time $P(R)$ is recorded. This is done by fitting a Gaussian to each distribution in the interval $4.71 \leq R/\text{Å} \leq 7.36$ $(5.33  \leq R/\text{Å} \leq 7.09)$ for $\mathrm{K_2}$ ($\mathrm{Rb_2}$) and extracting the center position from the fits. The time-dependence of $\langle R \rangle$, displayed by the white curves superimposed on $P(R,t)$ in \autoref{fig:Ak2_wp_im}, exhibits distinct periodic oscillations for the two alkali dimers during the entire 100~ps time window.

\begin{figure}[h!]
    \centering
    \includegraphics{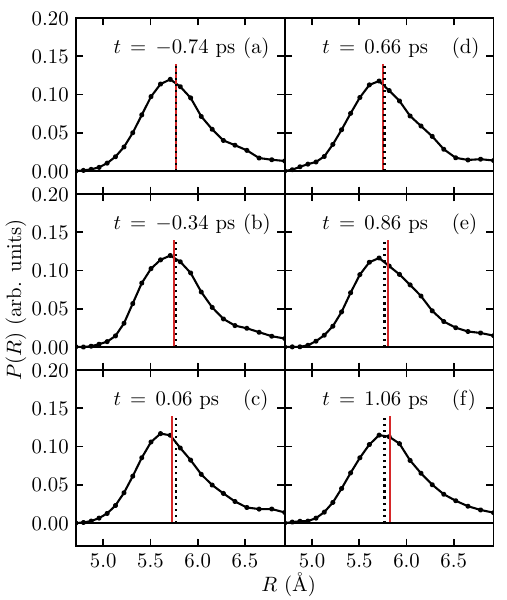}
    \caption{(a)--(f) Experimental $P(R)$ distributions (black dots) for \Kdimer at selected time-delays immediately before, during, and after interaction with the laser field (compare with the simulated distributions in \autoref{fig:dse}(a1)--(a6)). The vertical dotted line shown in each panel indicates the position of the average value of $\langle R \rangle (t)$ for $t<-0.6$~ps, denoted by $\overline{\langle R \rangle}$ (see text). The vertical red line indicates the expectation value $\langle R \rangle (t)$ for the distribution shown in each panel. Distributions are normalized to $\sum_i P(R_i)=1$.}
    \label{fig:wp}
\end{figure}


Before analyzing the spectral content of $\langle R \rangle (t)$, we focus on $P(R,t)$ in the interval corresponding to just before, during, and directly after the pump pulse. \Autoref{fig:wp} shows $P(R)$ for $\mathrm{K_2}$ at the same selected times as shown in \autoref{fig:dse}(a1)--(a6). The vertical dotted lines in \autoref{fig:wp} indicate the position of the mean value of $\langle R \rangle$ denoted as $\overline{\langle R \rangle}$, determined from the $R$-distributions recorded before the pump pulse arrives, where we anticipate $\langle R \rangle$ to be independent of time. In this sense, $\overline{\langle R \rangle}$ is equivalent to $\langle R \rangle_{v=0}$, which is annotated similarly in \autoref{fig:dse}. We find $\overline{\langle R \rangle}$~=~5.77~\AA, i.e., the same value as for $\langle R \rangle_{v=0}$. This similarity confirms the ability of the Coulomb explosion method to accurately capture bond distances of alkali dimers. The dotted lines serve as references for the displacements of the experimental $R$-distributions. On each panel in \autoref{fig:wp}, $\langle R \rangle$ is shown by the vertical red line.

At $t = -0.74$~ps, i.e., just before the pump pulse arrives, $\langle R \rangle$ is essentially equal to $\overline{\langle R \rangle}$ (per definition). At $t = -0.34$~ps, $P(R)$ is slightly shifted to lower $R$-values and, correspondingly,  $\langle R \rangle$ is slightly smaller than $\overline{\langle R \rangle}$. At $t = 0.06$~ps, the shift of $P(R)$ to smaller $R$-values increases, which is reflected by the vertical red line moving further to the left. At longer times, shown in panels (d)--(f), $P(R)$ moves gradually to larger $R$-values, and, correspondingly, the red line moves gradually to the right of the dotted black line. The time dependence of $P(R)$ and of $\langle R \rangle$, shown at the six times in \autoref{fig:wp}, agrees qualitatively well with that of the calculated $\left| \Psi(R,t) \right|^2$ in \autoref{fig:dse}. This agreement supports our interpretation of the mechanism that leads to the formation of the vibrational wave packet, as outlined in \autoref{sec:dse}. The magnitude of the observed displacements of the experimental $P(R)$ distributions are considerably smaller than those of the simulated dynamics as evident by comparison of \autoref{fig:wp} with \autoref{fig:dse}. The main reason for the observed discrepancy between the vibrational amplitudes in the experiment and the simulation is, we believe, that incoherent averaging of the experimental observables due to the spatial intensity distribution of the pump pulse is not included in the simulations.


In comparison with the theoretical distributions, the FWHM of the experimental $P(R)$ distributions reported in this work are larger by roughly a factor of $2$ and $2.5$ for $\mathrm{K_2}$ and $\mathrm{Rb_2}$, respectively. This is expected based on similar observations of Coulomb explosion imaging of alkali dimers in the vibrational ground level of either the \singlet state or the \triplet state~\cite{kristensen_laser-induced_2023,albrechtsen_laser-induced_2024}. We believe the reasons for the experimental broadening are mainly internuclear movement during the laser pulse and the energy resolution of the VMI spectrometer, see discussion in Ref.~\cite{kristensen_laser-induced_2023,albrechtsen_laser-induced_2024}.

\subsection{Spectral analysis}
\begin{figure}[h!]
    \centering
    \includegraphics[width=.47\textwidth]{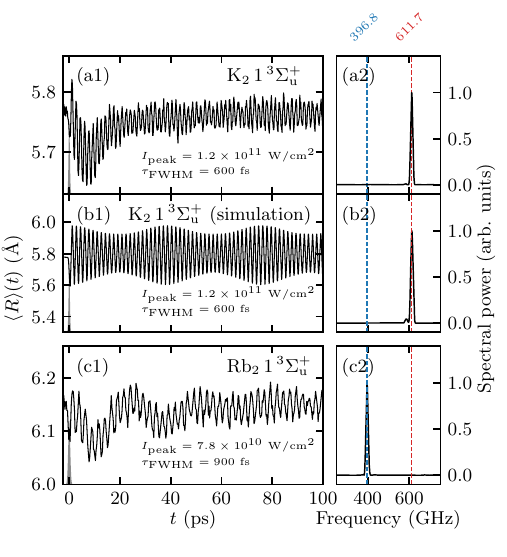}
    \caption{(a1) and (c1): $\langle R \rangle(t)$ for the $\mathrm{K_2}$ and $\mathrm{Rb_2}$ measurements, respectively. (a2) and (c2): The corresponding power spectrum of $\langle R \rangle(t)$. (b1) and (b2): Simulated results for \ce{K2} obtained with the model described in \autoref{sec:dse}. The vertical dashed lines in (a2)--(c2) indicate the central position of the main spectral peaks with the value annotated at the top.  The intensity profiles of the respective pump pulses are depicted as the grey shaded areas bounded by Gaussians with $\mathrm{FWHM} = \tau_{\mathrm{pump}}$. The listed parameters pertain to the pump pulses.}
    \label{fig:Ak2_R_FFT}
\end{figure}

\Autoref{fig:Ak2_R_FFT}(a1) and (c1) redisplay $\langle R \rangle (t)$ for $\mathrm{K_2}$ and $\mathrm{Rb_2}$ with a significantly expanded $y$-scale to make the oscillations more visible. The two $\langle R \rangle (t)$ traces are Fourier transformed to analyze their spectral content. Each power spectrum, displayed in \autoref{fig:Ak2_R_FFT}(a2) and (c2), is dominated by a single intense peak. Gaussian fits are used to find the central positions of the peaks, giving $611.7$~GHz for \Kdimer and $396.8$~GHz for \Rbdimer corresponding to a vibrational period of $1.63$~ps and $2.52$~ps, respectively. The FWHM of the spectral peaks is $\sim 13$~GHz, which is mainly determined by the 100~ps delay range.

As a comparison, we calculated $\nu_{1,0}$, defined by \autoref{eq:freqs}, for isolated \Kdimer and \Rbdimer in the \triplet state. In the \Kdimer measurement, we detect \ce{^39K+} ions. Based on the relative abundance of the potassium isotopes, $93.3\%$ of the \ce{^39K+} ions should originate from the \ce{^39K2} isotopologue and $6.7\%$ from the \ce{^39K^41K} isotopologue. For \ce{^39K2}, the calculation yields $611.8$~GHz and $604.7$~GHz for \ce{^39K^41K}~\footnote{We found the vibrational energy levels by solving the stationary vibrational Schrodinger equation based on the internuclear potentials given in~\cite{bauer_accurate_2019}.}. The similarity of $\nu_{1,0}$ for \ce{^39K2} to the frequency of the dominant peak in the spectrum, \autoref{fig:Ak2_R_FFT}(a2), identifies the periodic oscillations in $\langle R \rangle(t)$, \autoref{fig:Ak2_R_FFT}(a1), as originating from the quantum beat (coherence) between the $v$~=~0 and 1 vibrational states, i.e., the pump pulse has created a vibrational wave packet in \Kdimer composed mainly of the $v$~=~0 and 1 vibrational states. We note that the current spectral resolution does not allow the peak from \ce{^39K^41K}, with an expected central position $\sim$~604.1~GHz, to be distinguished from the main peak from \ce{^39K2}, in particular not since the amplitude of the \ce{^39K^41K} peak is a factor of 14 smaller than the amplitude of the \ce{^39K2} peak.

In the \Rbdimer measurement, we detect \ce{^85Rb+} ions. Here, the relative isotope abundances mean that $72.2\%$ of the \ce{^85Rb+} ions come from the \ce{^85Rb2} isotopologue and $27.8\%$ from the \ce{^85Rb^87Rb} isotopologue. The calculation yields $\nu_{1,0} = 398.3$ GHz~for \ce{^85Rb2} and $396.1$~GHz for \ce{^85Rb^87Rb}. The close resemblance of these two frequencies to the frequency of the strong peak in the spectrum, \autoref{fig:Ak2_R_FFT}(c2), allows us to conclude that the pump pulse created a vibrational wave packet in the \ce{Rb} dimers, which also primarily is comprised of the $v$~=~0 and 1 vibrational states. These results are consistent with the observations reported in Ref.~\cite{gruner_vibrational_2011} where a fs pump laser pulse, centered at 800 nm, created a coherent superposition of mainly the $v$ = 0 and 1 vibrational levels in the \triplet state of \Rbdimer via resonant stimulated Raman scattering.

In the spectrum for \Kdimer, an additional, small peak centered at $583.0$ GHz is visible. Within the spectral resolution of the measurement, this frequency is sufficiently close to the theoretical value of $\nu_{2,1}$ = $585.3$ GHz that we identify it is as originating from the quantum beat between the $v$ = 1 and 2 vibrational states. These experimental findings are fully consistent with the theoretical model outlined in \autoref{sec:dse}, i.e., the polarizability interaction produces a wave packet composed of the $v$ = 0, 1 and 2 levels in the \triplet state. Notably, for the duration and intensity of the pump laser pulse, the wave packet is mainly consisting of the two lowest vibrational states.


We observe a pronounced transient modulation of $\langle R \rangle (t)$ for \ce{K2} during the first $\sim 30$ ps, \autoref{fig:Ak2_R_FFT}(a1). A similar transient feature is also present for \ce{Rb2} during the first $\sim 70$~ps as seen in \autoref{fig:Ak2_R_FFT}(c1). Recently, we showed that a pump pulse with similar parameters ($\lambda=1.30~\mu$m, $\tau = 1$ ps, $I_\mathrm{peak} = 5.0 \times 10^{10}$ $\mathrm{W/cm^2}$), induces rotational wave packets in the dimers~\cite{kranabetter_nonadiabatic_2023}. In the current work, the pump pulse also induces alignment and we observe that the time-dependent degree of alignment of the internuclear axis with respect to the laser polarization, quantified by \cost$(t)$, is inversely correlated with the modulation of $\langle R \rangle(t)$. This suggests the presence of a probe effect dependent on the dimer alignment as the explanation for the observed transient modulations in $\langle R \rangle (t)$. One potential explanation is the distortion of the \triplet potential curve by the strong probe pulse. This distortion will cause the bond distance to shorten and the effect will be largest when the dimer axis points along the polarization axis of the probe field due to the $\theta$-dependence in \autoref{eq:pol_int_full}. Thus, the effect should correlate directly with the degree of alignment, which is consistent with what we observe.

Finally, \autoref{fig:Ak2_R_FFT}(b1) and (b2) present the simulated results for \ce{K2}. The agreement between the measured and calculated $\langle R \rangle (t)$ is very good in terms of the oscillatory structure. This is even more evident in the spectra, where both the main peak, corresponding to the $v$ = 0--1 coherence and the minor peak, corresponding to the $v$ = 1--2 correspondence, fall at the same central values. The comparison between panels (a1) and (b1) also shows that the amplitude of the simulated $\langle R \rangle (t)$ is larger than that of the experimental trace. We ascribe this to the fact that the simulations include neither rotation of the dimers nor averaging over the different intensities within the volume of the probe laser focus. In addition, the early modulation of the measured $\langle R \rangle (t)$ is not present in the calculated $\langle R \rangle (t)$ since the probe process is not taken into account in the simulations.


\subsection{Long-term dynamics: Wave packet dephasing}

\begin{figure*}[t]
    \centering
    \includegraphics[width=1.0\textwidth]{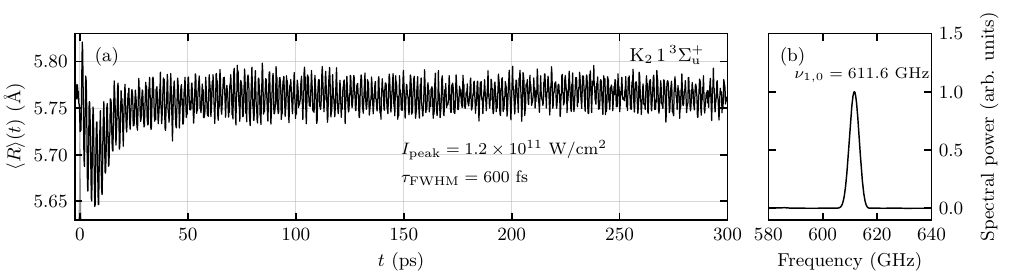}
    \caption{(a) $\langle R\rangle(t)$ for the $\mathrm{K_2}$ measurement in the time range out to 300~ps. The data for the first 100~ps are identical to those shown in \autoref{fig:Ak2_R_FFT}(a1). The grey shape depicts the intensity profile of the pump pulse. The listed parameters are for the pump pulse. (b) Power spectrum of $\langle R\rangle(t)$.}
    \label{fig:K2_wide}
\end{figure*}

\Autoref{fig:K2_wide}(a) shows $\langle R\rangle(t)$ for \Kdimer out to $300$~ps, a time interval covering $\sim 180$ vibrational periods. The corresponding power spectrum, displayed in \autoref{fig:K2_wide}(b), is dominated by a single intense peak similar to the spectrum obtained for the first $100$~ps, \autoref{fig:Ak2_R_FFT}(a2). The central frequency, $611.6$~GHz, is essentially the same as for the peak in \autoref{fig:Ak2_R_FFT}(a2), whereas the FWHM, $4$~GHz, is a factor of three smaller, reflecting that the data were recorded for three times as long. Inspection of \autoref{fig:K2_wide}(a) shows that the amplitude of $\langle R\rangle(t)$ gradually decreases. During the first 10--20~ps, the amplitude is 0.035~\AA, while it is reduced to 0.02~\AA{} towards the end of the scan. As a reference for these amplitude oscillations, we estimate the statistical uncertainty on the $\langle R\rangle$-measurements. This was done by determining the standard deviation $\sigma$ on $\langle R\rangle$ using the data recorded for $t < -0.7$~ps, i.e., before the pump pulse, where $\langle R\rangle$ should be time-independent. We find $\sigma$~=~0.007~\AA.

\begin{figure}[h]
    \centering
    \includegraphics{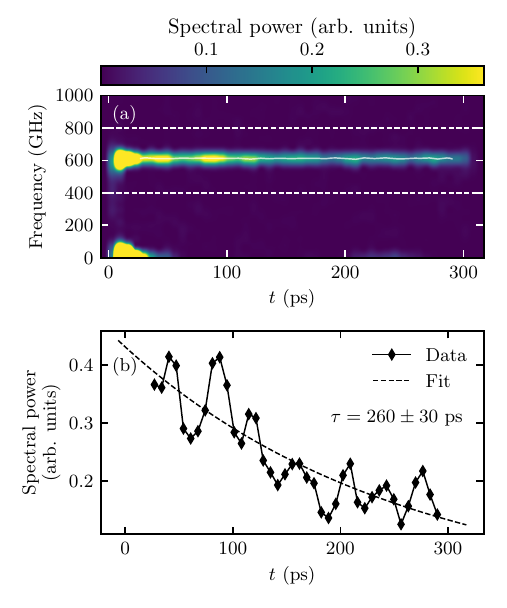}
    \caption{(a) Spectrogram obtained from sliding Fourier transform (SFT) analysis of $\langle R \rangle (t)$ for \Kdimer, see text. The dashed horizontal lines mark the region of interest. (b) Black diamonds: Spectral power of the signal at $612$~GHZ as a function of time. The full line just connects the points. Dashed line: Exponential fit used to determine the decay time $\tau$ of the oscillation amplitude in $\langle R\rangle(t)$. }
    \label{fig:K2_stft}
\end{figure}

To further analyze the temporal evolution of the amplitude of $\langle R\rangle(t)$, we employ a sliding Fourier transform (SFT). This approach sacrifices some spectral resolution to gain temporal resolution. The SFT is obtained by partitioning the $\langle R\rangle(t)$ data into overlapping segments, each 20~ps in length, with their midpoints, $t_i$, spaced 7~ps apart, and then Fourier transforming each segment. A Gaussian tapering function with a FWHM of $10$~ps was used on each segment before the Fourier transformation. The resulting power spectra are put together, using interpolation, to yield the
spectrogram, $|\mathrm{SFT}(t,\nu)|^2$, shown in \autoref{fig:K2_stft}(a). Two prominent features are visible. First, the low-frequency feature, which is only present for $t\leq 50$~ps, is attributed to the probe effect correlated with the alignment dynamics. Second, the horizontal line centered at around 612~GHz is identified as the dominant frequency of $\langle R\rangle(t)$. The signal gradually weakens reflecting the gradual decrease of the amplitude of $\langle R\rangle(t)$.


To quantify the decrease of the amplitude of $\langle R\rangle(t)$ we fit Gaussians to the power spectra $|\mathrm{SFT}(t=t_i,\nu)|^2$, i.e., to each vertical slice of the spectrogram at the midpoints of the time segments. Only the region between the two dashed horizontal lines is used. From this, we extract the central frequency, plotted as a white curve on the spectrogram in \autoref{fig:K2_stft}(a), as well as the spectral power, \autoref{fig:K2_stft}(b). While the former remains constant at 612~GHz, the latter exhibits a decay with a periodic modulation. The period of the oscillations is $\sim 40$~ps corresponding to a frequency of $25$~GHz, which is very close to the frequency difference between $\nu_{1,0}$ and $\nu_{2,1}$. We analyze the decay by fitting an exponential decay function $\propto \exp(-t/\tau)$ to the data points excluding points for $t\leq 40$~ps because they are affected by the alignment-dependent probe effect. The result is $\tau = 260 \pm 30$~ps, where the uncertainty is the $1\sigma$ standard deviation derived from the fit. We interpret $\tau$ as the lifetime of the coherent terms that contribute to the spectral power at 612~GHz. In addition to the 1-0 coherence with frequency $\nu_{1,0}$ it is also the 2-1 coherence with frequency $\nu_{2,1}$ because within the spectral resolution, $\sim 50$~GHz, defined by the $20$~ps time segments, we cannot formally distinguish between these two contributions. However, given that the Fourier transformation of the full time interval showed that the 1-0 coherence is the dominating term, we interpret $\tau$ as the lifetime of the 1-0 coherence. The 260~ps lifetime is close to the 0.3~ns lifetime found for similar vibrational wave packets in the \triplet state of \Rbdimer explored by Gr\"{u}ner et al.~\cite{gruner_vibrational_2011}. Their detailed comparison of the experimental results with dissipative quantum simulations of the wave packet dynamics revealed that the 0.3~ns lifetime -- giving rise to the gradual decay of the oscillations they observed in the time-dependent ionization yield of the vibrating dimer -- was caused by population decay of the excited vibrational states due to the interaction with the nearby helium droplet~\footnote{We note that these lifetimes are comparable to lifetimes calculated for the $v$ = 1 level in ground state diatomic molecules inside He nanodroplet, see Ref.~\cite{blancafort-jorquera_vibrational_2021}, but more than an order of magnitude longer than the decoherence lifetime measured for vibrational wave packets in an electronically excited state for In dimers also in the droplet interior, see Ref.~\cite{thaler_long-lived_2020}. }. We did not perform similar quantum simulations since the detailed influence of the weak dimer-droplet coupling is not the central theme of our work. However, the similarity of the lifetime observed for \Kdimer to that of \Rbdimer in Ref.~\cite{gruner_vibrational_2011} indicates that dissipation of vibrational energy to the droplet is also the mechanism that causes the gradual decay of $\langle R\rangle(t)$ although we cannot exclude that it is also due to pure dephasing.

Finally, we note that it has previously been discussed that alkali dimers may desorb from the helium droplet following electronic excitation~\cite{claas_wave_2006}. For \Rbdimer and \Kdimer studied here, the vibrational wave packets are created in the lowest-lying triplet state, i.e., no electronic excitation is induced, so do the dimers desorb in this case? For both dimers, we observed a gradual decay of the oscillations in $\langle R\rangle(t)$ during the 300~ps (100~ps) time window explored. Thus, for this time the dimers continue to interact with the droplet, i.e., they remain attached. This is in line with the conclusion by Gr\"{u}ner et al. on the vibrational wave packets formed in the lowest-lying triplet state of \Rbdimer~\cite{gruner_vibrational_2011}. The non-desorption is corroborated by our recent experimental observations of distinct, non-gas-phase time-dependent alignment dynamics for \Nadimer, \Kdimer, and \Rbdimer in either the \singlet or the \triplet state. Thus, we conclude that alkali dimers initially residing at the surface of helium droplets remain there after they are set into vibration or rotation by a non-resonant laser pulse through the polarizability interaction~\cite{kranabetter_nonadiabatic_2023,kristensen_alignment_2024}.

\section{Comparison to previous works and outlook}
In previous studies of the vibrational motion of alkali dimers on He droplet surfaces, the wave packets were probed by single ionization, induced through multiphoton absorption from the pump pulse, and measured by recording the dimer cation or electron yields as a function of time~\cite{claas_wave_2006,claas_wave_2007,mudrich_spectroscopy_2009,schlesinger_dissipative_2010, gruner_vibrational_2011}. This protocol can capture the dynamics of the wave packets, i.e. vibrational frequencies and possible dispersion and subsequent revivals, similar to what can be obtained with the Coulomb explosion probe method. Also, in terms of time resolution and sensitivity the two methods operate similarly. The key advantage of the Coulomb explosion method is its ability to provide direct structural information about the dimers, something which is not possible by measuring time-dependent ionization yields.

Here, we showed that it can image the nuclear structure during vibrational motion of \ce{K2} and \ce{Rb2} in their lowest-lying triplet state. Coulomb explosion should perform similarly for vibration of the three other homonuclear alkali dimers \ce{Li2}, \ce{Na2}, and \ce{Cs2}, and to any of the heteronuclear dimers~\cite{albrechtsen_laser-induced_2024}, not just in the lowest-lying triplet state but also in the singlet ground state. For the latter, shorter ($\sim 100$~fs) pump pulses will be needed due to the larger vibrational constants of these more tightly bound systems. Other classes of molecules where the Coulomb explosion method may be rewarding include dimers of alkaline earth metal atoms or dimers of one alkali atom and one alkaline earth atom~\cite{lackner_helium-droplet-assisted_2014}. Only a few spectroscopic studies of these systems have been reported, so very little of their structure and dynamics on helium droplets is known. With Coulomb explosion, it should be possible to measure both equilibrium bond distances, internuclear wavefunctions, vibrational and rotational dynamics and if the dimers are located on the surface or immersed inside of the droplet. To ensure that the electrostatic fragmentation of the dimers, induced by the probe pulse, is characterized by a pure Coulombic interaction, it will, probably, be necessary to remove both of the two valence electrons from each alkaline earth atom. In practice, this should be straightforward since the second ionization potential of any of the alkaline earth atoms is much lower than that of a helium atom.

An interesting extension of the current work is, we believe, to apply timed Coulomb explosion to a wave packet created in the electronic ground state of an alkali dimer cation, for instance \ce{K2+}. This should be possible by single photon ionization of \Kdimer in the \triplet state. A cation interacts much stronger with the helium environment than the corresponding neutral system does as demonstrated recently for the time-resolved solvation of \ce{Na+} ions in helium droplets~\cite{albrechtsen_observing_2023}. Thus, we expect that the interaction between \ce{K2+} and the droplet will lead to both pure dephasing and vibrational relaxation much faster than for the neutral dimer case, possibly on a time scale not much longer than the vibrational period. The ability of the Coulomb explosion to measure the distribution of bond distances could allow direct observation of the vibrational relaxation ending, probably, by cooling of \ce{K2+} to the $v$~=~0 state.

\section{Conclusion}

In conclusion, we demonstrated that the periodic, time-dependent distribution of internuclear distances resulting from a vibrational wave packet formed in the \triplet state of \Kdimer and \Rbdimer on a helium droplet surface can be measured with femtosecond-timed Coulomb explosion. Fourier analysis of $\langle R\rangle(t)$ showed that the wave packets are predominantly composed of the $v$~=~0 and $v$~=~1 vibrational states. This two-state wave packet is the reason for the cosine-like oscillations observed in $\langle R\rangle(t)$. The experimental findings are in good qualitative agreement with our numerical modeling based on the dynamic Stark effect as the mechanism for the creation of the wave packets. In the case of \Kdimer, the oscillations in $\langle R\rangle(t)$ persisted for $>180$ periods but their amplitude gradually decreased from 0.035~\AA{} to 0.020~\AA{} on a time scale of $\sim$~0.3~ns. In analogy with previous work on vibrational wave packet dynamics in \Rbdimer triplet states, we ascribe this decrease to vibrational relaxation of the dimers due to the interaction with the nearby droplet.

\section*{Acknowledgments}
H.S. acknowledges support from The Villum Foundation through Villum Investigator Grant No. 25886. We also thank Jan Thøgersen for carefully maintaining the laser system.

\section*{Author Declarations}
The authors have no conflicts to disclose.

\section*{Data Availability Statement}
The data that support the findings of this study are available from the corresponding author upon reasonable request.


%

\end{document}